\hsize=31pc 
\vsize=49pc 
\lineskip=0pt 
\parskip=0pt plus 1pt 
\hfuzz=1pt   
\vfuzz=2pt 
\pretolerance=2500 
\tolerance=5000 
\vbadness=5000 
\hbadness=5000 
\widowpenalty=500 
\clubpenalty=200 
\brokenpenalty=500 
\predisplaypenalty=200 
\voffset=-1pc 
\nopagenumbers      
\catcode`@=11 
\newif\ifams 
\amsfalse 
%
%
%
\newfam\bdifam 
\newfam\bsyfam 
\newfam\bssfam 
\newfam\msafam 
\newfam\msbfam 
\newif\ifxxpt    
\newif\ifxviipt  
\newif\ifxivpt   
\newif\ifxiipt   
\newif\ifxipt    
\newif\ifxpt     
\newif\ifixpt    
\newif\ifviiipt  
\newif\ifviipt   
\newif\ifvipt    
\newif\ifvpt     
%
%
\def\headsize#1#2{\def\headb@seline{#2}%
                \ifnum#1=20\def\HEAD{twenty}%
                           \def\smHEAD{twelve}%
                           \def\vsHEAD{nine}%
                           \ifxxpt\else\xdef\f@ntsize{\HEAD}%
                           \def\m@g{4}\def\s@ze{20.74}%
                           \loadheadfonts\xxpttrue\fi 
                           \ifxiipt\else\xdef\f@ntsize{\smHEAD}%
                           \def\m@g{1}\def\s@ze{12}%
                           \loadxiiptfonts\xiipttrue\fi 
                           \ifixpt\else\xdef\f@ntsize{\vsHEAD}%
                           \def\s@ze{9}%
                           \loadsmallfonts\ixpttrue\fi 
                      \else 
                \ifnum#1=17\def\HEAD{seventeen}%
                           \def\smHEAD{eleven}%
                           \def\vsHEAD{eight}%
                           \ifxviipt\else\xdef\f@ntsize{\HEAD}%
                           \def\m@g{3}\def\s@ze{17.28}%
                           \loadheadfonts\xviipttrue\fi 
                           \ifxipt\else\xdef\f@ntsize{\smHEAD}%
                           \loadxiptfonts\xipttrue\fi 
                           \ifviiipt\else\xdef\f@ntsize{\vsHEAD}%
                           \def\s@ze{8}%
                           \loadsmallfonts\viiipttrue\fi 
                      \else\def\HEAD{fourteen}%
                           \def\smHEAD{ten}%
                           \def\vsHEAD{seven}%
                           \ifxivpt\else\xdef\f@ntsize{\HEAD}%
                           \def\m@g{2}\def\s@ze{14.4}%
                           \loadheadfonts\xivpttrue\fi 
                           \ifxpt\else\xdef\f@ntsize{\smHEAD}%
                           \def\s@ze{10}%
                           \loadxptfonts\xpttrue\fi 
                           \ifviipt\else\xdef\f@ntsize{\vsHEAD}%
                           \def\s@ze{7}%
                           \loadviiptfonts\viipttrue\fi 
                \ifnum#1=14\else 
                \message{Header size should be 20, 17 or 14 point 
                              will now default to 14pt}\fi 
                \fi\fi\headfonts} 
%
%
\def\textsize#1#2{\def\textb@seline{#2}%
                 \ifnum#1=12\def\TEXT{twelve}%
                           \def\smTEXT{eight}%
                           \def\vsTEXT{six}%
                           \ifxiipt\else\xdef\f@ntsize{\TEXT}%
                           \def\m@g{1}\def\s@ze{12}%
                           \loadxiiptfonts\xiipttrue\fi 
                           \ifviiipt\else\xdef\f@ntsize{\smTEXT}%
                           \def\s@ze{8}%
                           \loadsmallfonts\viiipttrue\fi 
                           \ifvipt\else\xdef\f@ntsize{\vsTEXT}%
                           \def\s@ze{6}%
                           \loadviptfonts\vipttrue\fi 
                      \else 
                \ifnum#1=11\def\TEXT{eleven}%
                           \def\smTEXT{seven}%
                           \def\vsTEXT{five}%
                           \ifxipt\else\xdef\f@ntsize{\TEXT}%
                           \def\s@ze{11}%
                           \loadxiptfonts\xipttrue\fi 
                           \ifviipt\else\xdef\f@ntsize{\smTEXT}%
                           \loadviiptfonts\viipttrue\fi 
                           \ifvpt\else\xdef\f@ntsize{\vsTEXT}%
                           \def\s@ze{5}%
                           \loadvptfonts\vpttrue\fi 
                      \else\def\TEXT{ten}%
                           \def\smTEXT{seven}%
                           \def\vsTEXT{five}%
                           \ifxpt\else\xdef\f@ntsize{\TEXT}%
                           \loadxptfonts\xpttrue\fi 
                           \ifviipt\else\xdef\f@ntsize{\smTEXT}%
                           \def\s@ze{7}%
                           \loadviiptfonts\viipttrue\fi 
                           \ifvpt\else\xdef\f@ntsize{\vsTEXT}%
                           \def\s@ze{5}%
                           \loadvptfonts\vpttrue\fi 
                \ifnum#1=10\else 
                \message{Text size should be 12, 11 or 10 point 
                              will now default to 10pt}\fi 
                \fi\fi\textfonts} 
%
%
\def\smallsize#1#2{\def\smallb@seline{#2}%
                 \ifnum#1=10\def\SMALL{ten}%
                           \def\smSMALL{seven}%
                           \def\vsSMALL{five}%
                           \ifxpt\else\xdef\f@ntsize{\SMALL}%
                           \loadxptfonts\xpttrue\fi 
                           \ifviipt\else\xdef\f@ntsize{\smSMALL}%
                           \def\s@ze{7}%
                           \loadviiptfonts\viipttrue\fi 
                           \ifvpt\else\xdef\f@ntsize{\vsSMALL}%
                           \def\s@ze{5}%
                           \loadvptfonts\vpttrue\fi 
                       \else 
                 \ifnum#1=9\def\SMALL{nine}%
                           \def\smSMALL{six}%
                           \def\vsSMALL{five}%
                           \ifixpt\else\xdef\f@ntsize{\SMALL}%
                           \def\s@ze{9}%
                           \loadsmallfonts\ixpttrue\fi 
                           \ifvipt\else\xdef\f@ntsize{\smSMALL}%
                           \def\s@ze{6}%
                           \loadviptfonts\vipttrue\fi 
                           \ifvpt\else\xdef\f@ntsize{\vsSMALL}%
                           \def\s@ze{5}%
                           \loadvptfonts\vpttrue\fi 
                       \else 
                           \def\SMALL{eight}%
                           \def\smSMALL{six}%
                           \def\vsSMALL{five}%
                           \ifviiipt\else\xdef\f@ntsize{\SMALL}%
                           \def\s@ze{8}%
                           \loadsmallfonts\viiipttrue\fi 
                           \ifvipt\else\xdef\f@ntsize{\smSMALL}%
                           \def\s@ze{6}%
                           \loadviptfonts\vipttrue\fi 
                           \ifvpt\else\xdef\f@ntsize{\vsSMALL}%
                           \def\s@ze{5}%
                           \loadvptfonts\vpttrue\fi 
                 \ifnum#1=8\else\message{Small size should be 10, 9 or  
                            8 point will now default to 8pt}\fi 
                \fi\fi\smallfonts} 
\def\F@nt{\expandafter\font\csname} 
\def\Sk@w{\expandafter\skewchar\csname} 
\def\@nd{\endcsname} 
\def\@step#1{ scaled \magstep#1} 
\def\@half{ scaled \magstephalf} 
\def\@t#1{ at #1pt} 
%
%
\def\loadheadfonts{\bigf@nts 
\F@nt \f@ntsize bdi\@nd=cmmib10 \@t{\s@ze}%
\Sk@w \f@ntsize bdi\@nd='177 
\F@nt \f@ntsize bsy\@nd=cmbsy10 \@t{\s@ze}%
\Sk@w \f@ntsize bsy\@nd='60 
\F@nt \f@ntsize bss\@nd=cmssbx10 \@t{\s@ze}} 
%
%
\def\loadxiiptfonts{\bigf@nts 
\F@nt \f@ntsize bdi\@nd=cmmib10 \@step{\m@g}%
\Sk@w \f@ntsize bdi\@nd='177 
\F@nt \f@ntsize bsy\@nd=cmbsy10 \@step{\m@g}%
\Sk@w \f@ntsize bsy\@nd='60 
\F@nt \f@ntsize bss\@nd=cmssbx10 \@step{\m@g}} 
%
%
\def\loadxiptfonts{%
\font\elevenrm=cmr10 \@half 
\font\eleveni=cmmi10 \@half 
\skewchar\eleveni='177 
\font\elevensy=cmsy10 \@half 
\skewchar\elevensy='60 
\font\elevenex=cmex10 \@half 
\font\elevenit=cmti10 \@half 
\font\elevensl=cmsl10 \@half 
\font\elevenbf=cmbx10 \@half 
\font\eleventt=cmtt10 \@half 
\ifams\font\elevenmsa=msam10 \@half 
\font\elevenmsb=msbm10 \@half\else\fi 
\font\elevenbdi=cmmib10 \@half 
\skewchar\elevenbdi='177 
\font\elevenbsy=cmbsy10 \@half 
\skewchar\elevenbsy='60 
\font\elevenbss=cmssbx10 \@half} 
%
%
\def\loadxptfonts{%
\font\tenbdi=cmmib10 
\skewchar\tenbdi='177 
\font\tenbsy=cmbsy10  
\skewchar\tenbsy='60 
\ifams\font\tenmsa=msam10  
\font\tenmsb=msbm10\else\fi 
\font\tenbss=cmssbx10}%
%
%
\def\loadsmallfonts{\smallf@nts 
\ifams 
\F@nt \f@ntsize ex\@nd=cmex\s@ze 
\else 
\F@nt \f@ntsize ex\@nd=cmex10\fi 
\F@nt \f@ntsize it\@nd=cmti\s@ze 
\F@nt \f@ntsize sl\@nd=cmsl\s@ze 
\F@nt \f@ntsize tt\@nd=cmtt\s@ze} 
%
%
\def\loadviiptfonts{%
\font\sevenit=cmti7 
\font\sevensl=cmsl8 at 7pt 
\ifams\font\sevenmsa=msam7  
\font\sevenmsb=msbm7 
\font\sevenex=cmex7 
\font\sevenbsy=cmbsy7 
\font\sevenbdi=cmmib7\else 
\font\sevenex=cmex10 
\font\sevenbsy=cmbsy10 at 7pt 
\font\sevenbdi=cmmib10 at 7pt\fi 
\skewchar\sevenbsy='60 
\skewchar\sevenbdi='177 
\font\sevenbss=cmssbx10 at 7pt}%
%
%
\def\loadviptfonts{\smallf@nts 
\ifams\font\sixex=cmex7 at 6pt\else 
\font\sixex=cmex10\fi 
\font\sixit=cmti7 at 6pt} 
%
%
\def\loadvptfonts{%
\font\fiveit=cmti7 at 5pt 
\ifams\font\fiveex=cmex7 at 5pt 
\font\fivebdi=cmmib5 
\font\fivebsy=cmbsy5 
\font\fivemsa=msam5  
\font\fivemsb=msbm5\else 
\font\fiveex=cmex10 
\font\fivebdi=cmmib10 at 5pt 
\font\fivebsy=cmbsy10 at 5pt\fi 
\skewchar\fivebdi='177 
\skewchar\fivebsy='60 
\font\fivebss=cmssbx10 at 5pt} 
\def\bigf@nts{%
\F@nt \f@ntsize rm\@nd=cmr10 \@step{\m@g}%
\F@nt \f@ntsize i\@nd=cmmi10 \@step{\m@g}%
\Sk@w \f@ntsize i\@nd='177 
\F@nt \f@ntsize sy\@nd=cmsy10 \@step{\m@g}%
\Sk@w \f@ntsize sy\@nd='60 
\F@nt \f@ntsize ex\@nd=cmex10 \@step{\m@g}%
\F@nt \f@ntsize it\@nd=cmti10 \@step{\m@g}%
\F@nt \f@ntsize sl\@nd=cmsl10 \@step{\m@g}%
\F@nt \f@ntsize bf\@nd=cmbx10 \@step{\m@g}%
\F@nt \f@ntsize tt\@nd=cmtt10 \@step{\m@g}%
\ifams 
\F@nt \f@ntsize msa\@nd=msam10 \@step{\m@g}%
\F@nt \f@ntsize msb\@nd=msbm10 \@step{\m@g}\else\fi} 
\def\smallf@nts{%
\F@nt \f@ntsize rm\@nd=cmr\s@ze 
\F@nt \f@ntsize i\@nd=cmmi\s@ze  
\Sk@w \f@ntsize i\@nd='177 
\F@nt \f@ntsize sy\@nd=cmsy\s@ze 
\Sk@w \f@ntsize sy\@nd='60 
\F@nt \f@ntsize bf\@nd=cmbx\s@ze  
\ifams 
\F@nt \f@ntsize bdi\@nd=cmmib\s@ze  
\F@nt \f@ntsize bsy\@nd=cmbsy\s@ze  
\F@nt \f@ntsize msa\@nd=msam\s@ze  
\F@nt \f@ntsize msb\@nd=msbm\s@ze 
\else 
\F@nt \f@ntsize bdi\@nd=cmmib10 \@t{\s@ze}%
\F@nt \f@ntsize bsy\@nd=cmbsy10 \@t{\s@ze}\fi  
\Sk@w \f@ntsize bdi\@nd='177 
\Sk@w \f@ntsize bsy\@nd='60 
\F@nt \f@ntsize bss\@nd=cmssbx10 \@t{\s@ze}}%
%
%
\def\headfonts{%
\textfont0=\csname\HEAD rm\@nd         
\scriptfont0=\csname\smHEAD rm\@nd 
\scriptscriptfont0=\csname\vsHEAD rm\@nd 
\def\rm{\fam0\csname\HEAD rm\@nd 
\def\sc{\csname\smHEAD rm\@nd}}%
\textfont1=\csname\HEAD i\@nd          
\scriptfont1=\csname\smHEAD i\@nd 
\scriptscriptfont1=\csname\vsHEAD i\@nd 
\textfont2=\csname\HEAD sy\@nd         
\scriptfont2=\csname\smHEAD sy\@nd 
\scriptscriptfont2=\csname\vsHEAD sy\@nd 
\textfont3=\csname\HEAD ex\@nd         
\scriptfont3=\csname\smHEAD ex\@nd 
\scriptscriptfont3=\csname\smHEAD ex\@nd 
\textfont\itfam=\csname\HEAD it\@nd    
\scriptfont\itfam=\csname\smHEAD it\@nd 
\scriptscriptfont\itfam=\csname\vsHEAD it\@nd 
\def\it{\fam\itfam\csname\HEAD it\@nd 
\def\sc{\csname\smHEAD it\@nd}}%
\textfont\slfam=\csname\HEAD sl\@nd    
\def\sl{\fam\slfam\csname\HEAD sl\@nd 
\def\sc{\csname\smHEAD sl\@nd}}%
\textfont\bffam=\csname\HEAD bf\@nd    
\scriptfont\bffam=\csname\smHEAD bf\@nd 
\scriptscriptfont\bffam=\csname\vsHEAD bf\@nd 
\def\bf{\fam\bffam\csname\HEAD bf\@nd 
\def\sc{\csname\smHEAD bf\@nd}}%
\textfont\ttfam=\csname\HEAD tt\@nd    
\def\tt{\fam\ttfam\csname\HEAD tt\@nd}%
\textfont\bdifam=\csname\HEAD bdi\@nd  
\scriptfont\bdifam=\csname\smHEAD bdi\@nd 
\scriptscriptfont\bdifam=\csname\vsHEAD bdi\@nd 
\def\bdi{\fam\bdifam\csname\HEAD bdi\@nd}%
\textfont\bsyfam=\csname\HEAD bsy\@nd  
\scriptfont\bsyfam=\csname\smHEAD bsy\@nd 
\def\bsy{\fam\bsyfam\csname\HEAD bsy\@nd}%
\textfont\bssfam=\csname\HEAD bss\@nd  
\scriptfont\bssfam=\csname\smHEAD bss\@nd 
\scriptscriptfont\bssfam=\csname\vsHEAD bss\@nd 
\def\bss{\fam\bssfam\csname\HEAD bss\@nd}%
\ifams 
\textfont\msafam=\csname\HEAD msa\@nd  
\scriptfont\msafam=\csname\smHEAD msa\@nd 
\scriptscriptfont\msafam=\csname\vsHEAD msa\@nd 
\textfont\msbfam=\csname\HEAD msb\@nd  
\scriptfont\msbfam=\csname\smHEAD msb\@nd 
\scriptscriptfont\msbfam=\csname\vsHEAD msb\@nd 
\else\fi 
\normalbaselineskip=\headb@seline pt%
\setbox\strutbox=\hbox{\vrule height.7\normalbaselineskip  
depth.3\baselineskip width0pt}%
\def\sc{\csname\smHEAD rm\@nd}\normalbaselines\bf} 
%
%
\def\textfonts{%
\textfont0=\csname\TEXT rm\@nd         
\scriptfont0=\csname\smTEXT rm\@nd 
\scriptscriptfont0=\csname\vsTEXT rm\@nd 
\def\rm{\fam0\csname\TEXT rm\@nd 
\def\sc{\csname\smTEXT rm\@nd}}%
\textfont1=\csname\TEXT i\@nd          
\scriptfont1=\csname\smTEXT i\@nd 
\scriptscriptfont1=\csname\vsTEXT i\@nd 
\textfont2=\csname\TEXT sy\@nd         
\scriptfont2=\csname\smTEXT sy\@nd 
\scriptscriptfont2=\csname\vsTEXT sy\@nd 
\textfont3=\csname\TEXT ex\@nd         
\scriptfont3=\csname\smTEXT ex\@nd 
\scriptscriptfont3=\csname\smTEXT ex\@nd 
\textfont\itfam=\csname\TEXT it\@nd    
\scriptfont\itfam=\csname\smTEXT it\@nd 
\scriptscriptfont\itfam=\csname\vsTEXT it\@nd 
\def\it{\fam\itfam\csname\TEXT it\@nd 
\def\sc{\csname\smTEXT it\@nd}}%
\textfont\slfam=\csname\TEXT sl\@nd    
\def\sl{\fam\slfam\csname\TEXT sl\@nd 
\def\sc{\csname\smTEXT sl\@nd}}%
\textfont\bffam=\csname\TEXT bf\@nd    
\scriptfont\bffam=\csname\smTEXT bf\@nd 
\scriptscriptfont\bffam=\csname\vsTEXT bf\@nd 
\def\bf{\fam\bffam\csname\TEXT bf\@nd 
\def\sc{\csname\smTEXT bf\@nd}}%
\textfont\ttfam=\csname\TEXT tt\@nd    
\def\tt{\fam\ttfam\csname\TEXT tt\@nd}%
\textfont\bdifam=\csname\TEXT bdi\@nd  
\scriptfont\bdifam=\csname\smTEXT bdi\@nd 
\scriptscriptfont\bdifam=\csname\vsTEXT bdi\@nd 
\def\bdi{\fam\bdifam\csname\TEXT bdi\@nd}%
\textfont\bsyfam=\csname\TEXT bsy\@nd  
\scriptfont\bsyfam=\csname\smTEXT bsy\@nd 
\def\bsy{\fam\bsyfam\csname\TEXT bsy\@nd}%
\textfont\bssfam=\csname\TEXT bss\@nd  
\scriptfont\bssfam=\csname\smTEXT bss\@nd 
\scriptscriptfont\bssfam=\csname\vsTEXT bss\@nd 
\def\bss{\fam\bssfam\csname\TEXT bss\@nd}%
\ifams 
\textfont\msafam=\csname\TEXT msa\@nd  
\scriptfont\msafam=\csname\smTEXT msa\@nd 
\scriptscriptfont\msafam=\csname\vsTEXT msa\@nd 
\textfont\msbfam=\csname\TEXT msb\@nd  
\scriptfont\msbfam=\csname\smTEXT msb\@nd 
\scriptscriptfont\msbfam=\csname\vsTEXT msb\@nd 
\else\fi 
\normalbaselineskip=\textb@seline pt 
\setbox\strutbox=\hbox{\vrule height.7\normalbaselineskip  
depth.3\baselineskip width0pt}%
\everymath{}%
\def\sc{\csname\smTEXT rm\@nd}\normalbaselines\rm} 
%
%
\def\smallfonts{%
\textfont0=\csname\SMALL rm\@nd         
\scriptfont0=\csname\smSMALL rm\@nd 
\scriptscriptfont0=\csname\vsSMALL rm\@nd 
\def\rm{\fam0\csname\SMALL rm\@nd 
\def\sc{\csname\smSMALL rm\@nd}}%
\textfont1=\csname\SMALL i\@nd          
\scriptfont1=\csname\smSMALL i\@nd 
\scriptscriptfont1=\csname\vsSMALL i\@nd 
\textfont2=\csname\SMALL sy\@nd         
\scriptfont2=\csname\smSMALL sy\@nd 
\scriptscriptfont2=\csname\vsSMALL sy\@nd 
\textfont3=\csname\SMALL ex\@nd         
\scriptfont3=\csname\smSMALL ex\@nd 
\scriptscriptfont3=\csname\smSMALL ex\@nd 
\textfont\itfam=\csname\SMALL it\@nd    
\scriptfont\itfam=\csname\smSMALL it\@nd 
\scriptscriptfont\itfam=\csname\vsSMALL it\@nd 
\def\it{\fam\itfam\csname\SMALL it\@nd 
\def\sc{\csname\smSMALL it\@nd}}%
\textfont\slfam=\csname\SMALL sl\@nd    
\def\sl{\fam\slfam\csname\SMALL sl\@nd 
\def\sc{\csname\smSMALL sl\@nd}}%
\textfont\bffam=\csname\SMALL bf\@nd    
\scriptfont\bffam=\csname\smSMALL bf\@nd 
\scriptscriptfont\bffam=\csname\vsSMALL bf\@nd 
\def\bf{\fam\bffam\csname\SMALL bf\@nd 
\def\sc{\csname\smSMALL bf\@nd}}%
\textfont\ttfam=\csname\SMALL tt\@nd    
\def\tt{\fam\ttfam\csname\SMALL tt\@nd}%
\textfont\bdifam=\csname\SMALL bdi\@nd  
\scriptfont\bdifam=\csname\smSMALL bdi\@nd 
\scriptscriptfont\bdifam=\csname\vsSMALL bdi\@nd 
\def\bdi{\fam\bdifam\csname\SMALL bdi\@nd}%
\textfont\bsyfam=\csname\SMALL bsy\@nd  
\scriptfont\bsyfam=\csname\smSMALL bsy\@nd 
\def\bsy{\fam\bsyfam\csname\SMALL bsy\@nd}%
\textfont\bssfam=\csname\SMALL bss\@nd  
\scriptfont\bssfam=\csname\smSMALL bss\@nd 
\scriptscriptfont\bssfam=\csname\vsSMALL bss\@nd 
\def\bss{\fam\bssfam\csname\SMALL bss\@nd}%
\ifams 
\textfont\msafam=\csname\SMALL msa\@nd  
\scriptfont\msafam=\csname\smSMALL msa\@nd 
\scriptscriptfont\msafam=\csname\vsSMALL msa\@nd 
\textfont\msbfam=\csname\SMALL msb\@nd  
\scriptfont\msbfam=\csname\smSMALL msb\@nd 
\scriptscriptfont\msbfam=\csname\vsSMALL msb\@nd 
\else\fi 
\normalbaselineskip=\smallb@seline pt%
\setbox\strutbox=\hbox{\vrule height.7\normalbaselineskip  
depth.3\baselineskip width0pt}%
\everymath{}%
\def\sc{\csname\smSMALL rm\@nd}\normalbaselines\rm}%
\everydisplay{\indenteddisplay 
   \gdef\labeltype{\eqlabel}}%
%
%
\def\hexnumber@#1{\ifcase#1 0\or 1\or 2\or 3\or 4\or 5\or 6\or 7\or 8\or 
 9\or A\or B\or C\or D\or E\or F\fi} 
\edef\bffam@{\hexnumber@\bffam} 
\edef\bdifam@{\hexnumber@\bdifam} 
\edef\bsyfam@{\hexnumber@\bsyfam} 
\def\undefine#1{\let#1\undefined} 
\def\newsymbol#1#2#3#4#5{\let\next@\relax 
 \ifnum#2=\thr@@\let\next@\bdifam@\else 
 \ifams 
 \ifnum#2=\@ne\let\next@\msafam@\else 
 \ifnum#2=\tw@\let\next@\msbfam@\fi\fi 
 \fi\fi 
 \mathchardef#1="#3\next@#4#5} 
\def\mathhexbox@#1#2#3{\relax 
 \ifmmode\mathpalette{}{\m@th\mathchar"#1#2#3}%
 \else\leavevmode\hbox{$\m@th\mathchar"#1#2#3$}\fi} 

\def\bi#1{{\fam\bdifam\relax#1}} 
%
%
\ifams\input amsmacro\fi 
%
%
\newsymbol\bitGamma 3000 
\newsymbol\bitDelta 3001 
\newsymbol\bitTheta 3002 
\newsymbol\bitLambda 3003 
\newsymbol\bitXi 3004 
\newsymbol\bitPi 3005 
\newsymbol\bitSigma 3006 
\newsymbol\bitUpsilon 3007 
\newsymbol\bitPhi 3008 
\newsymbol\bitPsi 3009 
\newsymbol\bitOmega 300A 
\newsymbol\balpha 300B 
\newsymbol\bbeta 300C 
\newsymbol\bgamma 300D 
\newsymbol\bdelta 300E 
\newsymbol\bepsilon 300F 
\newsymbol\bzeta 3010 
\newsymbol\bfeta 3011 
\newsymbol\btheta 3012 
\newsymbol\biota 3013 
\newsymbol\bkappa 3014 
\newsymbol\blambda 3015 
\newsymbol\bmu 3016 
\newsymbol\bnu 3017 
\newsymbol\bxi 3018 
\newsymbol\bpi 3019 
\newsymbol\brho 301A 
\newsymbol\bsigma 301B 
\newsymbol\btau 301C 
\newsymbol\bupsilon 301D 
\newsymbol\bphi 301E 
\newsymbol\bchi 301F 
\newsymbol\bpsi 3020 
\newsymbol\bomega 3021 
\newsymbol\bvarepsilon 3022 
\newsymbol\bvartheta 3023 
\newsymbol\bvaromega 3024 
\newsymbol\bvarrho 3025 
\newsymbol\bvarzeta 3026 
\newsymbol\bvarphi 3027 
\newsymbol\bpartial 3040 
\newsymbol\bell 3060 
\newsymbol\bimath 307B 
\newsymbol\bjmath 307C 
\mathchardef\binfty "0\bsyfam@31 
\mathchardef\bnabla "0\bsyfam@72 
\mathchardef\bdot "2\bsyfam@01 
\mathchardef\bGamma "0\bffam@00 
\mathchardef\bDelta "0\bffam@01 
\mathchardef\bTheta "0\bffam@02 
\mathchardef\bLambda "0\bffam@03 
\mathchardef\bXi "0\bffam@04 
\mathchardef\bPi "0\bffam@05 
\mathchardef\bSigma "0\bffam@06 
\mathchardef\bUpsilon "0\bffam@07 
\mathchardef\bPhi "0\bffam@08 
\mathchardef\bPsi "0\bffam@09 
\mathchardef\bOmega "0\bffam@0A 
\mathchardef\itGamma "0100 
\mathchardef\itDelta "0101 
\mathchardef\itTheta "0102 
\mathchardef\itLambda "0103 
\mathchardef\itXi "0104 
\mathchardef\itPi "0105 
\mathchardef\itSigma "0106 
\mathchardef\itUpsilon "0107 
\mathchardef\itPhi "0108 
\mathchardef\itPsi "0109 
\mathchardef\itOmega "010A 
\mathchardef\Gamma "0000 
\mathchardef\Delta "0001 
\mathchardef\Theta "0002 
\mathchardef\Lambda "0003 
\mathchardef\Xi "0004 
\mathchardef\Pi "0005 
\mathchardef\Sigma "0006 
\mathchardef\Upsilon "0007 
\mathchardef\Phi "0008 
\mathchardef\Psi "0009 
\mathchardef\Omega "000A 
%
%
\newcount\firstpage  \firstpage=1  
\newcount\jnl                      
\newcount\secno                    
\newcount\subno                    
\newcount\subsubno                 
\newcount\appno                    
\newcount\tabno                    
\newcount\figno                    
\newcount\countno                  
\newcount\refno                    
\newcount\eqlett     \eqlett=97    
\newif\ifletter 
\newif\ifwide 
\newif\ifnotfull 
\newif\ifaligned 
\newif\ifnumbysec   
\newif\ifappendix 
\newif\ifnumapp 
\newif\ifssf 
\newif\ifppt 
\newdimen\t@bwidth 
\newdimen\c@pwidth 
\newdimen\digitwidth                    
\newdimen\argwidth                      
\newdimen\secindent    \secindent=5pc   
\newdimen\textind    \textind=16pt      
\newdimen\tempval                       
\newskip\beforesecskip 
\def\beforesecspace{\vskip\beforesecskip\relax} 
\newskip\beforesubskip 
\def\beforesubspace{\vskip\beforesubskip\relax} 
\newskip\beforesubsubskip 
\def\beforesubsubspace{\vskip\beforesubsubskip\relax} 
\newskip\secskip 
\def\secspace{\vskip\secskip\relax} 
\newskip\subskip 
\def\subspace{\vskip\subskip\relax} 
\newskip\insertskip 
\def\insertspace{\vskip\insertskip\relax} 
\def\sp@ce{\ifx\next*\let\next=\@ssf 
               \else\let\next=\@nossf\fi\next} 
\def\@ssf#1{\nobreak\secspace\global\ssftrue\nobreak} 
\def\@nossf{\nobreak\secspace\nobreak\noindent\ignorespaces} 
\def\subsp@ce{\ifx\next*\let\next=\@sssf 
               \else\let\next=\@nosssf\fi\next} 
\def\@sssf#1{\nobreak\subspace\global\ssftrue\nobreak} 
\def\@nosssf{\nobreak\subspace\nobreak\noindent\ignorespaces} 
\beforesecskip=24pt plus12pt minus8pt 
\beforesubskip=12pt plus6pt minus4pt 
\beforesubsubskip=12pt plus6pt minus4pt 
\secskip=12pt plus 2pt minus 2pt 
\subskip=6pt plus3pt minus2pt 
\insertskip=18pt plus6pt minus6pt%
\fontdimen16\tensy=2.7pt 
\fontdimen17\tensy=2.7pt 
%
%
\def\eqlabel{(\ifappendix\applett 
               \ifnumbysec\ifnum\secno>0 \the\secno\fi.\fi 
               \else\ifnumbysec\the\secno.\fi\fi\the\countno)} 
\def\seclabel{\ifappendix\ifnumapp\else\applett\fi 
    \ifnum\secno>0 \the\secno 
    \ifnumbysec\ifnum\subno>0.\the\subno\fi\fi\fi 
    \else\the\secno\fi\ifnum\subno>0.\the\subno 
         \ifnum\subsubno>0.\the\subsubno\fi\fi} 
\def\tablabel{\ifappendix\applett\fi\the\tabno} 
\def\figlabel{\ifappendix\applett\fi\the\figno} 
\def\gac{\global\advance\countno by 1} 
%
%
 
\def\vfootnote#1{\insert\footins\bgroup 
\interlinepenalty=\interfootnotelinepenalty 
\splittopskip=\ht\strutbox 
\splitmaxdepth=\dp\strutbox \floatingpenalty=20000 
\leftskip=0pt \rightskip=0pt \spaceskip=0pt \xspaceskip=0pt%
\noindent\smallfonts\rm #1\ \ignorespaces\footstrut\futurelet\next\fo@t} 
%
%
\def\endinsert{\egroup 
    \if@mid \dimen@=\ht0 \advance\dimen@ by\dp0 
       \advance\dimen@ by12\p@ \advance\dimen@ by\pagetotal 
       \ifdim\dimen@>\pagegoal \@midfalse\p@gefalse\fi\fi 
    \if@mid \insertspace \box0 \par \ifdim\lastskip<\insertskip 
    \removelastskip \penalty-200 \insertspace \fi 
    \else\insert\topins{\penalty100 
       \splittopskip=0pt \splitmaxdepth=\maxdimen  
       \floatingpenalty=0 
       \ifp@ge \dimen@=\dp0 
       \vbox to\vsize{\unvbox0 \kern-\dimen@}%
       \else\box0\nobreak\insertspace\fi}\fi\endgroup}    
%
%
%
\def\ind{\hbox to \secindent{\hfill}} 
%
%

%
%
 
%
%
\def\indeqn#1{\alignedfalse\displ@y\halign{\hbox to \displaywidth 
    {$\ind\@lign\displaystyle##\hfil$}\crcr #1\crcr}} 
%
%
\def\indalign#1{\alignedtrue\displ@y \tabskip=0pt  
  \halign to\displaywidth{\ind$\@lign\displaystyle{##}$\tabskip=0pt 
    &$\@lign\displaystyle{{}##}$\hfill\tabskip=\centering 
    &\llap{$\@lign\hbox{\rm##}$}\tabskip=0pt\crcr 
    #1\crcr}} 
\def\fl{{\hskip-\secindent}} 
\def\indenteddisplay#1$${\indispl@y{#1 }} 
\def\indispl@y#1{\disptest#1\eqalignno\eqalignno\disptest} 
\def\disptest#1\eqalignno#2\eqalignno#3\disptest{%
    \ifx#3\eqalignno 
    \indalign#2%
    \else\indeqn{#1}\fi$$} 
%
%
 
%
%
 
%
%
 
%
%
 
%
%

\def\ns{\noalign{\vskip-3pt}}

%
 
%
%
\def\bhbar{\rlap{\kern1pt\raise.4ex\hbox{\bf\char'40}}\bi{h}} 

\def\d{{\rm d}} 
\def\e{{\rm e}} 
 
\def\frac#1#2{{#1\over#2}} 
\ifams 
\def\lap{\lesssim} 
\def\gap{\gtrsim}

\else

\def\gap{\;\lower3pt\hbox{$\buildrel > \over \sim$}\;}%
\def\lap{\;\lower3pt\hbox{$\buildrel < \over \sim$}\;}\fi 
\def\i{{\rm i}} 
\chardef\ii="10 
\def\tqs{\hbox to 25pt{\hfil}}

\def\Bbbone{1\kern-.22em {\rm l}} 
%
%
\def\rp{\raise8pt\hbox{$\scriptstyle\prime$}} 
%
%
%
%

%
%
\def\[#1\]{\setbox0=\hbox{$\dsty#1$}\argwidth=\wd0 
    \setbox0=\hbox{$\left[\box0\right]$}\advance\argwidth by -\wd0 
    \left[\kern.3\argwidth\box0\kern.3\argwidth\right]} 
%
%
\def\lsb#1\rsb{\setbox0=\hbox{$#1$}\argwidth=\wd0 
    \setbox0=\hbox{$\left[\box0\right]$}\advance\argwidth by -\wd0 
    \left[\kern.3\argwidth\box0\kern.3\argwidth\right]} 
%
 
%
%
 
%
\def\pt(#1){({\it #1\/})} 
\let\dsty=\displaystyle

%
%
\def\reactions#1{\vskip 12pt plus2pt minus2pt%
\vbox{\hbox{\kern\secindent\vrule\kern12pt%
\vbox{\kern0.5pt\vbox{\hsize=24pc\parindent=0pt\smallfonts\rm NUCLEAR  
REACTIONS\strut\quad #1\strut}\kern0.5pt}\kern12pt\vrule}}} 
%
%
\def\slashchar#1{\setbox0=\hbox{$#1$}\dimen0=\wd0%
\setbox1=\hbox{/}\dimen1=\wd1%
\ifdim\dimen0>\dimen1%
\rlap{\hbox to \dimen0{\hfil/\hfil}}#1\else                                         
\rlap{\hbox to \dimen1{\hfil$#1$\hfil}}/\fi} 
%
%
\def\textindent#1{\noindent\hbox to \parindent{#1\hss}\ignorespaces} 
%
%
\def\opencirc{\raise1pt\hbox{$\scriptstyle{\bigcirc}$}} 
 
\ifams 
\def\opensqr{\hbox{$\square$}} 
 
\def\opentridown{\hbox{$\triangledown$}}

\else 
\def\opensqr{\vbox{\hrule height.4pt\hbox{\vrule width.4pt height3.5pt 
    \kern3.5pt\vrule width.4pt}\hrule height.4pt}} 
 
\def\opentridown{\raise1pt\hbox{$\scriptstyle\bigtriangledown$}}

\fi

%
%
\def\m@th{\mathsurround=0pt} 
%
%
\def\cases#1{%
\left\{\,\vcenter{\normalbaselines\openup1\jot\m@th%
     \ialign{$\displaystyle##\hfil$&\rm\tqs##\hfil\crcr#1\crcr}}\right.}%
%
%
\def\oldcases#1{\left\{\,\vcenter{\normalbaselines\m@th 
    \ialign{$##\hfil$&\rm\quad##\hfil\crcr#1\crcr}}\right.} 
%
%
\def\numcases#1{\left\{\,\vcenter{\baselineskip=15pt\m@th%
     \ialign{$\displaystyle##\hfil$&\rm\tqs##\hfil 
     \crcr#1\crcr}}\right.\hfill 
     \vcenter{\baselineskip=15pt\m@th%
     \ialign{\rlap{$\phantom{\displaystyle##\hfil}$}\tabskip=0pt&\en 
     \rlap{\phantom{##\hfil}}\crcr#1\crcr}}} 
\def\ptnumcases#1{\left\{\,\vcenter{\baselineskip=15pt\m@th%
     \ialign{$\displaystyle##\hfil$&\rm\tqs##\hfil 
     \crcr#1\crcr}}\right.\hfill 
     \vcenter{\baselineskip=15pt\m@th%
     \ialign{\rlap{$\phantom{\displaystyle##\hfil}$}\tabskip=0pt&\enpt 
     \rlap{\phantom{##\hfil}}\crcr#1\crcr}}\global\eqlett=97 
     \global\advance\countno by 1} 
%
%
\def\eq(#1){\ifaligned\@mp(#1)\else\hfill\llap{{\rm (#1)}}\fi} 
\def\ceq(#1){\ns\ns\ifaligned\@mp\fi\eq(#1)\cr\ns\ns} 
\def\eqpt(#1#2){\ifaligned\@mp(#1{\it #2\/}) 
                    \else\hfill\llap{{\rm (#1{\it #2\/})}}\fi} 
\let\eqno=\eq 
%
%
\countno=1 
 
\def\aleq{&\rm(\ifappendix\applett 
               \ifnumbysec\ifnum\secno>0 \the\secno\fi.\fi 
               \else\ifnumbysec\the\secno.\fi\fi\the\countno} 
\def\noaleq{\hfill\llap\bgroup\rm(\ifappendix\applett 
               \ifnumbysec\ifnum\secno>0 \the\secno\fi.\fi 
               \else\ifnumbysec\the\secno.\fi\fi\the\countno} 
\def\@mp{&} 
\def\en{\ifaligned\aleq)\else\noaleq)\egroup\fi\gac} 
\def\cen{\ns\ns\ifaligned\@mp\fi\en\cr\ns\ns} 
\def\enpt{\ifaligned\aleq{\it\char\the\eqlett})\else 
    \noaleq{\it\char\the\eqlett})\egroup\fi 
    \global\advance\eqlett by 1} 
\def\endpt{\ifaligned\aleq{\it\char\the\eqlett})\else 
    \noaleq{\it\char\the\eqlett})\egroup\fi 
    \global\eqlett=97\gac} 
%
%

\def\JPA{{\it J. Phys. A: Math. Gen.}} 

 

%
%

\def\APNY{{\it Ann. Phys., NY\/}}

\def\PR{{\it Phys. Rev.}} 
\def\PRL{{\it Phys. Rev. Lett.}}

\def\RMP{{\it Rev. Mod. Phys.}}

\def\ZP{{\it Z. Phys.}} 
\headline={\ifodd\pageno{\ifnum\pageno=\firstpage\hfill 
   \else\rrhead\fi}\else\lrhead\fi} 
\def\rrhead{\textfonts\hskip\secindent\it 
    \shorttitle\hfill\rm\folio} 
\def\lrhead{\textfonts\hbox to\secindent{\rm\folio\hss}%
    \it\aunames\hss} 
\footline={\ifnum\pageno=\firstpage \hfill\textfonts\rm\folio\fi} 
\def\@rticle#1#2{\vglue.5pc 
    {\parindent=\secindent \bf #1\par} 
     \vskip2.5pc 
    {\exhyphenpenalty=10000\hyphenpenalty=10000 
     \baselineskip=18pt\raggedright\noindent 
     \headfonts\bf#2\par}\futurelet\next\sh@rttitle}%
\def\title#1{\gdef\shorttitle{#1} 
    \vglue4pc{\exhyphenpenalty=10000\hyphenpenalty=10000  
    \baselineskip=18pt  
    \raggedright\parindent=0pt 
    \headfonts\bf#1\par}\futurelet\next\sh@rttitle}  

\def\article#1#2{\gdef\shorttitle{#2}\@rticle{#1}{#2}}  
\def\review#1{\gdef\shorttitle{#1}%
    \@rticle{REVIEW \ifpbm\else ARTICLE\fi}{#1}} 
\def\topical#1{\gdef\shorttitle{#1}%
    \@rticle{TOPICAL REVIEW}{#1}} 
\def\comment#1{\gdef\shorttitle{#1}%
    \@rticle{COMMENT}{#1}} 
\def\note#1{\gdef\shorttitle{#1}%
    \@rticle{NOTE}{#1}} 
\def\prelim#1{\gdef\shorttitle{#1}%
    \@rticle{PRELIMINARY COMMUNICATION}{#1}} 
\def\letter#1{\gdef\shorttitle{Letter to the Editor}%
     \gdef\aunames{Letter to the Editor} 
     \global\lettertrue\ifnum\jnl=7\global\letterfalse\fi 
     \@rticle{LETTER TO THE EDITOR}{#1}} 
\def\sh@rttitle{\ifx\next[\let\next=\sh@rt 
                \else\let\next=\f@ll\fi\next} 
\def\sh@rt[#1]{\gdef\shorttitle{#1}} 
\def\f@ll{} 
\def\author#1{\ifletter\else\gdef\aunames{#1}\fi\vskip1.5pc 
    {\parindent=\secindent   
     \hang\textfonts   
     \ifppt\bf\else\rm\fi#1\par}   
     \ifppt\bigskip\else\smallskip\fi 
     \futurelet\next\@unames} 
\def\@unames{\ifx\next[\let\next=\short@uthor 
                 \else\let\next=\@uthor\fi\next} 
\def\short@uthor[#1]{\gdef\aunames{#1}} 
\def\@uthor{} 
\def\address#1{{\parindent=\secindent 
    \exhyphenpenalty=10000\hyphenpenalty=10000 
\ifppt\textfonts\else\smallfonts\fi\hang\raggedright\rm#1\par}%
    \ifppt\bigskip\fi} 
\def\jl#1{\global\jnl=#1} 
\jl{0}%
\def\journal{\ifnum\jnl=1 J. Phys.\ A: Math.\ Gen.\  
        \else\ifnum\jnl=2 J. Phys.\ B: At.\ Mol.\ Opt.\ Phys.\  
        \else\ifnum\jnl=3 J. Phys.:\ Condens. Matter\  
        \else\ifnum\jnl=4 J. Phys.\ G: Nucl.\ Part.\ Phys.\  
        \else\ifnum\jnl=5 Inverse Problems\  
        \else\ifnum\jnl=6 Class. Quantum Grav.\  
        \else\ifnum\jnl=7 Network\  
        \else\ifnum\jnl=8 Nonlinearity\ 
        \else\ifnum\jnl=9 Quantum Opt.\ 
        \else\ifnum\jnl=10 Waves in Random Media\ 
        \else\ifnum\jnl=11 Pure Appl. Opt.\  
        \else\ifnum\jnl=12 Phys. Med. Biol.\ 
        \else\ifnum\jnl=13 Modelling Simulation Mater.\ Sci.\ Eng.\  
        \else\ifnum\jnl=14 Plasma Phys. Control. Fusion\  
        \else\ifnum\jnl=15 Physiol. Meas.\  
        \else\ifnum\jnl=16 Sov.\ Lightwave Commun.\ 
        \else\ifnum\jnl=17 J. Phys.\ D: Appl.\ Phys.\ 
        \else\ifnum\jnl=18 Supercond.\ Sci.\ Technol.\ 
        \else\ifnum\jnl=19 Semicond.\ Sci.\ Technol.\ 
        \else\ifnum\jnl=20 Nanotechnology\ 
        \else\ifnum\jnl=21 Meas.\ Sci.\ Technol.\  
        \else\ifnum\jnl=22 Plasma Sources Sci.\ Technol.\  
        \else\ifnum\jnl=23 Smart Mater.\ Struct.\  
        \else\ifnum\jnl=24 J.\ Micromech.\ Microeng.\ 
   \else Institute of Physics Publishing\  
   \fi\fi\fi\fi\fi\fi\fi\fi\fi\fi\fi\fi\fi\fi\fi 
   \fi\fi\fi\fi\fi\fi\fi\fi\fi} 
\let\abs=\beginabstract 

\let\endabs=\endabstract 
\def\submitted{\ifppt\noindent\textfonts\rm Submitted to \journal\par 
     \bigskip\fi} 
\def\today{\number\day\ \ifcase\month\or 
     January\or February\or March\or April\or May\or June\or 
     July\or August\or September\or October\or November\or 
     December\fi\space \number\year} 
\def\date{\ifppt\noindent\textfonts\rm  
     Date: \today\par\goodbreak\bigskip\fi} 
%
%
\def\pacs#1{\ifppt\noindent\textfonts\rm  
     PACS number(s): #1\par\bigskip\fi} 
%
 
%
%
\def\section#1{\ifppt\ifnum\secno=0\eject\fi\fi 
    \subno=0\subsubno=0\global\advance\secno by 1 
    \gdef\labeltype{\seclabel}\ifnumbysec\countno=1\fi 
    \goodbreak\beforesecspace\nobreak 
    \noindent{\bf \the\secno. #1}\par\futurelet\next\sp@ce} 
\def\subsection#1{\subsubno=0\global\advance\subno by 1 
     \gdef\labeltype{\seclabel}%
     \ifssf\else\goodbreak\beforesubspace\fi 
     \global\ssffalse\nobreak 
     \noindent{\it \the\secno.\the\subno. #1\par}%
     \futurelet\next\subsp@ce} 
\def\subsubsection#1{\global\advance\subsubno by 1 
     \gdef\labeltype{\seclabel}%
     \ifssf\else\goodbreak\beforesubsubspace\fi 
     \global\ssffalse\nobreak 
     \noindent{\it \the\secno.\the\subno.\the\subsubno. #1}\null.  
     \ignorespaces} 
%
 
%
%
\def\numappendix#1{\ifappendix\ifnumbysec\countno=1\fi\else 
    \countno=1\figno=0\tabno=0\fi 
    \subno=0\global\advance\appno by 1 
    \secno=\appno\gdef\applett{A}\gdef\labeltype{\seclabel}%
    \global\appendixtrue\global\numapptrue 
    \goodbreak\beforesecspace\nobreak 
    \noindent{\bf Appendix \the\appno. #1\par}%
    \futurelet\next\sp@ce} 
\def\numsubappendix#1{\global\advance\subno by 1\subsubno=0 
    \gdef\labeltype{\seclabel}%
    \ifssf\else\goodbreak\beforesubspace\fi 
    \global\ssffalse\nobreak 
    \noindent{\it A\the\appno.\the\subno. #1\par}%
    \futurelet\next\subsp@ce} 
\def\@ppendix#1#2#3{\countno=1\subno=0\subsubno=0\secno=0\figno=0\tabno=0 
    \gdef\applett{#1}\gdef\labeltype{\seclabel}\global\appendixtrue 
    \goodbreak\beforesecspace\nobreak 
    \noindent{\bf Appendix#2#3\par}\futurelet\next\sp@ce} 
\def\Appendix#1{\@ppendix{A}{. }{#1}} 
\def\appendix#1#2{\@ppendix{#1}{ #1. }{#2}} 
\def\App#1{\@ppendix{A}{ }{#1}} 
\def\app{\@ppendix{A}{}{}} 
\def\subappendix#1#2{\global\advance\subno by 1\subsubno=0 
    \gdef\labeltype{\seclabel}%
    \ifssf\else\goodbreak\beforesubspace\fi 
    \global\ssffalse\nobreak 
    \noindent{\it #1\the\subno. #2\par}%
    \nobreak\subspace\noindent\ignorespaces} 
%
%
\def\@ck#1{\ifletter\bigskip\noindent\ignorespaces\else 
    \goodbreak\beforesecspace\nobreak 
    \noindent{\bf Acknowledgment#1\par}%
    \nobreak\secspace\noindent\ignorespaces\fi} 
\def\ack{\@ck{s}} 
\def\ackn{\@ck{}} 
\def\n@ip#1{\goodbreak\beforesecspace\nobreak 
    \noindent\smallfonts{\it #1}. \rm\ignorespaces} 
\def\naip{\n@ip{Note added in proof}} 
\def\na{\n@ip{Note added}} 
 
%
%
 
%
 
%
%
 
%
 
%
 
\def\tablecont{\topinsert\global\advance\tabno by -1 
    \tablecaption{(continued)}} 
\def\tablecaption#1{\gdef\labeltype{\tablabel}\global\widefalse 
    \leftskip=\secindent\parindent=0pt 
    \global\advance\tabno by 1 
    \smallfonts{\bf Table \ifappendix\applett\fi\the\tabno.} \rm #1\par 
    \smallskip\futurelet\next\t@b} 
\def\t@b{\ifx\next*\let\next=\widet@b 
             \else\ifx\next[\let\next=\fullwidet@b 
                      \else\let\next=\narrowt@b\fi\fi 
             \next} 
\def\widet@b#1{\global\widetrue\global\notfulltrue 
    \t@bwidth=\hsize\advance\t@bwidth by -\secindent}  
\def\fullwidet@b[#1]{\global\widetrue\global\notfullfalse 
    \leftskip=0pt\t@bwidth=\hsize}                   
\def\narrowt@b{\global\notfulltrue} 
\def\align{\catcode`?=13\ifnotfull\moveright\secindent\fi 
    \vbox\bgroup\halign\ifwide to \t@bwidth\fi 
    \bgroup\strut\tabskip=1.2pc plus1pc minus.5pc} 
\def\endalign{\egroup\egroup\catcode`?=12} 
 
%
%

%
%

%
 
%
%

%
 
\catcode`?=13 
\def\lineup{\setbox0=\hbox{\smallfonts\rm 0}%
    \digitwidth=\wd0%
    \def?{\kern\digitwidth}%
    \def\\{\hbox{$\phantom{-}$}}%
    \def\-{\llap{$-$}}} 
\catcode`?=12 
%
%
\def\sidetable#1#2{\hbox{\ifppt\hsize=18pc\t@bwidth=18pc 
                          \else\hsize=15pc\t@bwidth=15pc\fi 
    \parindent=0pt\vtop{\null #1\par}%
    \ifppt\hskip1.2pc\else\hskip1pc\fi 
    \vtop{\null #2\par}}}  
\def\lstable#1#2{\everypar{}\tempval=\hsize\hsize=\vsize 
    \vsize=\tempval\hoffset=-3pc 
    \global\tabno=#1\gdef\labeltype{\tablabel}%
    \noindent\smallfonts{\bf Table \ifappendix\applett\fi 
    \the\tabno.} \rm #2\par 
    \smallskip\futurelet\next\t@b} 
\def\inctabno{\global\advance\tabno by 1} 
%
%
 
%
 
%
\def\figure#1{\figc@ption{#1}\bigskip} 
\def\figc@ption#1{\global\advance\figno by 1\gdef\labeltype{\figlabel}%
   {\parindent=\secindent\smallfonts\hang 
    {\bf Figure \ifappendix\applett\fi\the\figno.} \rm #1\par}} 
%
%
\def\refHEAD{\goodbreak\beforesecspace 
     \noindent\textfonts{\bf References}\par 
     \let\ref=\rf 
     \nobreak\smallfonts\rm} 
\def\references{\refHEAD\parindent=0pt 
     \everypar{\hangindent=18pt\hangafter=1 
     \frenchspacing\rm}%
     \secspace} 
\def\rf#1{\par\noindent\hbox to 21pt{\hss #1\quad}\ignorespaces} 
%
 
%
 
%
%
\def\numrefjl#1#2#3#4#5{\par\rf{#1}#2 {\it #3 \bf #4} #5\par} 
%
%
\def\numrefbk#1#2#3#4{\par\rf{#1}#2 {\it #3} #4\par} 
%
%

%
%

%
\catcode`\@=12 
%
%
 
%
%
\def\jnlstyle{\pptfalse\headsize{14}{18}%
\textsize{10}{12}%
\smallsize{8}{10} 
\textind=16pt} 
%
%
 
%
%
 
%
\parindent=\textind 
%
\def\received#1{\insertspace 
     \parindent=\secindent\ifppt\textfonts\else\smallfonts\fi 
     \hang{Received #1}\rm } 
\def\appendix{\goodbreak\beforesecspace 
     \noindent\textfonts{\bf Appendix}\secspace} 
\headline={\ifodd\pageno{\ifnum\pageno=\firstpage\titlehead
   \else\rrhead\fi}\else\lrhead\fi} 
\def\lpsn#1#2{LPM-#1-LT#2}
 
\def\rrhead{\textfonts\hskip\secindent\it 
    \shorttitle\hfill\rm L\folio} 
\def\lrhead{\textfonts\hbox to\secindent{\rm L\folio\hss}%
   \it\aunames\hss} 
\footline={\ifnum\pageno=\firstpage
\smallfonts cond-mat/9612128\hfil\textfonts\rm L\folio\fi}   
\def\titlehead{\smallfonts J. Phys. A: Math. Gen. {\bf 30} (1997)
L105-111\hfil\lpsn{96}{2}} 

\firstpage=105
\pageno=105

\jnlstyle

\jl{1}
    
\overfullrule=0pt

\letter{Off-diagonal density profiles and conformal invariance}[Letter 
to the Editor]

\author{L~Turban\dag~and~F~Igl\'oi\ddag}[Letter to the Editor]

\address{\dag\  Laboratoire de Physique des 
Mat\'eriaux\footnote{\S}{Unit\'e de 
Recherche associ\'ee au CNRS no 155}, Universit\'e de Nancy~I, BP~239, 
F--54506~Vand\oe uvre l\`es Nancy Cedex, France}

\address{\ddag\  Reasearch Institute for Solid
State Physics, P.O. Box 49, H--1525~Budapest 114, Hungary 
and Institute for Theoretical Physics, Szeged University, Aradi V.
tere 1, H--6720~Szeged, Hungary}

\received{6 November 1996, in final form 18 December 1996}

\abs 
Off-diagonal profiles $\phi_{\rm od}(v)$ of 
local densities (e.g. order parameter
or energy density) are calculated at
the bulk critical point, by conformal
methods, on a strip with transverse coordinate 
$v$, for different types of
boundary conditions (free, fixed and mixed). Such profiles,
which are defined by the non-vanishing matrix element
$\langle0\vert\hat{\phi}(v)\vert\phi\rangle$ of the appropriate
operator $\hat{\phi}(v)$ between the ground state and the corresponding
lowest excited state of the strip Hamiltonian, enter into the expression of
two-point correlation functions on a strip. They are of interest in the
finite-size scaling study of bulk and 
surface critical behaviour since they
allow the elimination of regular contributions. 
The conformal profiles, which
are obtained through a conformal 
transformation of the correlation functions from
the half-plane to the strip, are 
in agreement with the results of a direct
calculation, for the energy density 
of the two-dimensional Ising model.
\endabs

\vglue1cm

\pacs{0550, 6842}

\submitted

\date

Following the pioneering work of Fisher and de Gennes~[1],
the study of order parameter and energy
density profiles near surfaces has been 
an active field of research during
the past years. These profiles have been calculated 
at, and near  the critical 
point, in the mean-field approximation~[2], using field-theoretical
approaches~[3] and through exact solutions~[4, 5].
Much progress has also been achieved in their calculation at bulk
criticality in two-dimensional ($2d$) systems making use of conformal
techniques~[6--12]. 

In a semi-infinite $2d$ system, the profile $\phi(y)$ of a fluctuating
quantity, such as the order parameter or the energy density, is obtained in the
transfer matrix formalism as the diagonal matrix element 
$\langle0\vert\hat{\phi}(y)\vert0\rangle$ of the corresponding operator 
$\hat{\phi}$ in the ground state $\vert0\rangle$ of the Hamiltonian
${\cal H}=-\ln{\cal T}$, where ${\cal T}$ denotes 
the transfer operator along 
the surface. 

On a strip of infinite length and finite 
width $L$, one may also consider the
off-diagonal profile $\phi_{\rm od}(v)$, where $v$ is the
transverse coordinate. The profile is then defined as the off-diagonal 
matrix element $\langle0\vert\hat{\phi}(v)
\vert\phi\rangle$ between the ground
state $\vert0\rangle$ of the Hamiltonian ${\cal H}$ on the strip and
its lowest excited state $\vert\phi\rangle$ 
leading to a nonvanishing matrix
element. 

These off-diagonal profiles are commonly used at the bulk critical point to
obtain informations about the surface 
and bulk critical behaviour via finite-size
scaling, while avoiding the regular terms 
which contribute to the diagonal
ones. Actually, in the  absence of 
an external symmetry-breaking field, an
off-diagonal matrix element has to be used 
to study the scaling behaviour of the
order parameter since the diagonal one vanishes, 
due to symmetry. They are also
of interest because of their high degree of universality. One may mention two
recent studies where off-diagonal profiles are considered at bulk criticality:
the spin 1/2--spin 1 Ising quantum chain in~[13] and the random
Ising quantum chain in~[14]. 

In the following, the scaling form of such off-diagonal matrix elements
is  obtained at the bulk critical point 
for two-dimensional conformally invariant
systems. It is deduced from the asymptotic 
behaviour of the appropriate two-point
correlation function in the half-space 
after a conformal transformation to the
strip geometry. We consider first  symmetric and then asymmetric boundary
conditions on the strip and compare our results to some exact expressions
of the energy-density profiles, obtained in the appendix, for the Ising model in
the extreme anisotropic limit. Since we use conformal methods noninvariant
boundary conditions, like a finite surface 
field for which an interesting scaling
behaviour has been recently observed~[15, 16], and
off-critical systems are here excluded. 
 
Let us first briefly review the Fisher-de Gennes result and its
consequence for the profile on a strip. We consider a nonvanishing
profile $\phi(y)$ on a semi-infinite $2d$ conformally invariant sytem at its
critical point with a surface at $y=0$. It may be the energy density profile
with any type of uniform boundary conditions 
or the order parameter profile with
fixed boundary conditions. The problem involves a single length scale, the
distance $y$ from the surface. Under a length rescaling  by a factor $b$,
the profile transforms  as $\phi(y/b)=b^{x_\phi}\phi(y)$ so that 
$$
\phi(y)={\cal A}y^{-x_\phi}
\eqno(1)
$$
where $x_\phi$ is the bulk scaling dimension of $\hat{\phi}$.
Now, making use of the conformal transformation $w=(L/\pi)\ln z$, with
$z=x+\i y=\rho\e^{\i\theta}$ and $w=u+iv$ one obtains
$$
\rho=\exp\left({\pi u\over L}\right)\; ,\qquad\theta={\pi v\over L}\; ,
\eqno(2)
$$
and the local dilatation factor is $b(z)=\vert{\d w/\d
z}\vert^{-1}=\pi\rho/L$. The half-plane $y>0$ is transformed into a
strip $-\infty<u<+\infty$, $0<v<L$ with the same boundary
conditions as the half-space on both edges~[17]. The profile
transforms  as~[18]:  
$$
\phi(w)=b(z)^{x_\phi}\phi(z)
\eqno(3)
$$
so that~[6]
$$
\phi(v)={\cal A}\left[{L\over\pi}\sin\left({\pi v\over
L}\right)\right]^{-x_\phi}\; .
\eqno(4)
$$

One may notice that the surface critical behaviour
is hidden in the sine variation. At a fixed distance $l\ll L$ from the
surface the profile behaves as  
$$ 
\phi(l)\simeq {\cal
A}l^{-x_\phi}\left(1+{\textstyle{1\over6}}\pi^2l^2x_\phi\, 
L^{-2}+\cdots\right)\; ,  
\eqno(5)
$$ 
in agreement with the Fisher-de Gennes conjecture which gives a
$L^{-d}$ correction term~[1] with an universal
amplitude~[9]. Generally, the exponent $d$ is expected to give the
scaling dimension $x_\phi^{\rm s}$ of $\hat{\phi}$ at any  surface transition
leading to a nonvanishing profile in the semi-infinite critical
system~[7,9,19--21]. The $O(N)$ model at the special transition (for $N<1$
in $2d$) provides a counter-example, the surface energy exponent being smaller
than $d$ in this
case~[22, 12, 23]\footnote{$\Vert$}{We thank E
Eisenriegler for pointing out this exception to us.}. 

Let us now turn to the calculation of the off-diagonal matrix element. On 
the critical semi-infinite system, 
conformal invariance strongly constrains the
form of the connected two-point correlation function, ${\cal
G}_{\phi\phi}^{\rm con}(x_1-x_2,y_1,y_2)$. Applying an
infinitesimal special conformal transformation which preserves the surface
geometry, one obtains a system of partial differential equations for ${\cal
G}_{\phi\phi}^{\rm con}$, from which the following scaling form is 
deduced~[24]:   
$$ \fl
{\cal G}_{\phi\phi}^{\rm con}(x_1-x_2,y_1,y_2)=(y_1y_2)^{-x_\phi}g(\omega)\;
,\qquad \omega={(x_1-x_2)^2+(y_1-y_2)^2\over y_1y_2}\; . 
\eqno(6)
$$
When $\omega\gg1$, ordinary scaling leads to
$g(\omega)\sim\omega^{-x_\phi^{\rm s}}$ where $x_\phi^{\rm s}$ is the scaling
dimension of $\hat{\phi}$ at the surface. This limit also corresponds to
$\rho_1\gg\rho_2$ where, using polar coordinates, (6)~may be
rewritten as:
$$
{\cal G}_{\phi\phi}^{\rm con}(\rho_1,\rho_2,\theta_1,\theta_2)\sim
(\rho_1\rho_2)^{-x_\phi}\left({\rho_1\over\rho_2}\right)^{-x_\phi^{\rm s}}
(\sin\theta_1\sin\theta_2)^{x_\phi^{\rm s}-x_\phi}\; .
\eqno(7)
$$
We now consider the connected two-point 
function at bulk criticality in the strip
geometry. In the same limit, it can be obtained through the conformal
transformation of~(7) from the half-plane to the strip geometry
with~[17]  
$$
{\cal G}_{\phi\phi}(w_1,w_2)=b(z_1)^{x_\phi}b(z_2)^{x_\phi}
{\cal G}_{\phi\phi}(z_1,z_2)\; , 
\eqno(8)
$$
giving:
$$\fl\eqalign{
{\cal G}_{\phi\phi}^{\rm con}(u_1-u_2,v_1,v_2)&=\left({\pi\over
L}\right)^{2x_\phi}(\rho_1\rho_2)^{x_\phi}
{\cal G}_{\phi\phi}^{\rm con}(\rho_1,\rho_2,\theta_1,\theta_2)\cr
&\sim\left({\pi\over L}\right)^{2x_\phi}\exp\left[-{\pi x_\phi^{\rm s}\over
L}(u_1-u_2)\right]\left[\sin\left({\pi v_1\over L}\right)
\sin\left({\pi v_2\over L}\right)\right]^{x_\phi^{\rm s}-x_\phi}.\cr}  
\eqno(9)
$$
On the strip, making use of the transfer operator $\e^{-\cal H}$, it can also
be written as an expansion over the eigenstates $\vert n\rangle$ of the
critical Hamiltonian ${\cal H}$ with eigenvalues $E_n$ and ground state energy
$E_0$: $$\fl
{\cal
G}_{\phi\phi}^{\rm con}(u_1-u_2,v_1,v_2)
=\sum_{n>0}\langle0\vert\hat{\phi}(v_1)\vert n\rangle\langle
n\vert\hat{\phi}(v_2)\vert 0\rangle\exp[-(E_n-E_0)(u_1-u_2)]\; .
\eqno(10)
$$ 
In the limit $\rho_1\gg\rho_2$ which corresponds to $u_1\gg u_2$ on the
strip, the sum is dominated by the contribution of the lowest exited state
$\vert\phi\rangle$ with a nonvanishing matrix element so that: 
$$\fl 
{\cal G}_{\phi\phi}^{\rm con}(u_1-u_2,v_1,v_2)
\simeq\langle0\vert\hat{\phi}(v_1)\vert
\phi\rangle\langle\phi\vert\hat{\phi}(v_2)\vert
0\rangle\exp[-(E_\phi-E_0)(u_1-u_2)]\; . 
\eqno(11)
$$   
Comparing with~(9) we recover the gap-exponent
relation $E_\phi-E_0=\pi x_\phi^{\rm s}/L$~[17] as a
by-product and we may identify the expression of the off-diagonal matrix
element 
$$
\phi_{\rm od}(v)\sim\left({L\over\pi}\right)^{-x_\phi}\left[\sin\left({\pi
v\over L}\right)\right]^{x_\phi^{\rm s}-x_\phi}\; ,
\eqno(12)
$$
with symmetric boundary conditions on the strip. 
The off-diagonal energy-density
profile for the Ising model given in equation~(A5) of the appendix, which is
valid for symmetric free or fixed boundary conditions, agrees with~(12) in
the continuum limit.

At a fixed distance $l\ll L$ from the surface,  $\phi_{\rm
od}(l)$, behaving as $L^{-x_\phi^{\rm s}}$, is quite appropriate to perform a
finite-size scaling study since the regular term which appeared in~(5) is
now avoided. One may notice that the diagonal profile $\phi(v)$ in~(4) is
formally recovered with $x_\phi^{\rm s}=0$ in~(12). Finally, in the
half-plane limit ($L\to\infty$), the amplitude of the off-diagonal profile
vanishes. This is consistent with the scaling considerations leading to
equation~(1): a nonvanishing profile on the semi-infinite system
necessarily gives the diagonal profile~(4) on the strip.
  
Next we consider the case of mixed boundary conditions on the half-space.
This means having different scale-invariant  boundary conditions $a$ and
$b$ on the positive and negative $x$-axes, respectively, with $a,b=f$ (free),
$+$ or $-$ (fixed). The plane-to-strip conformal transformation leads to
different boundary conditions $a$ and $b$ on opposite edges of the strip. The
method employed to determine the scaling behaviour of the two-point function
in~(6) can no longer be used since it supposed translation invariance along
the $x$-direction which is now broken by the mixed boundary conditions on the
surface. Fortunately, the two-point correlation functions with mixed boundary
conditions have been obtained explicitly for the Ising model using conformal
methods~[11]: $n$-point correlation functions in the half-space
with mixed boundary conditions are determined 
by the same differential equation
as a particular $2n+2$-point  bulk correlation function. Using the results of
Burkhardt and Xue~[11], when $\rho_1/\rho_2\gg1$, the asymptotic
spin-spin correlation functions read 
$$\fl\eqalign{ 
{\cal G}_{\sigma\sigma}^{+-}&\sim(\rho_1\rho_2\sin\theta_1\sin\theta_2)
^{-1/8}\left[\cos\theta_1\cos\theta_2+\sin^2\theta_1\sin^2\theta_2
\left({\rho_2\over\rho_1}\right)+\cdots\right]\cr 
{\cal G}_{\sigma\sigma}^{+f}&\sim(\rho_1\rho_2\sin\theta_1\sin\theta_2)
^{-1/8}(\cos\theta_1\cos\theta_2)^{1/2}\cr
&\qquad\qquad\qquad\qquad\times\Biggl[1+{1\over2}
\sin\theta_1\sin\theta_2\tan\left({\theta_1\over2}\right)
\tan\left({\theta_2\over2}\right)
{\rho_2\over\rho_1}+\cdots\Biggr]\; ,\cr} 
\eqno(13)
$$
whereas the following expressions are obtained for the energy-energy
correlation functions:
$$\fl\eqalign{
{\cal G}_{\varepsilon\varepsilon}^{+-}&\sim(\rho_1\rho_2
\sin\theta_1\sin\theta_2)^{-1}\Biggl[(1-4\sin^2\theta_1)(1-4\sin^2\theta_2)
+\Biggr.\cr
&\qquad\qquad\qquad\qquad\qquad\qquad\Biggl.+64\sin^2\theta_1\sin^2\theta_2
\cos\theta_1\cos\theta_2
\left({\rho_2\over\rho_1}\right)+\cdots\Biggr]\cr
{\cal
G}_{\varepsilon\varepsilon}^{+f}&\sim(\rho_1\rho_2\sin\theta_1\sin\theta_2)
^{-1}\left[\cos\theta_1\cos\theta_2+8\sin^2\theta_1\sin^2\theta_2
\left({\rho_2\over\rho_1}\right)+\cdots\right]\; .\cr}
\eqno(14)
$$
Making use of the conformal transformation to the strip geometry as in the
first line of equation~(9) and comparing to the expansion of the
correlation function on the strip with asymmetric boundary conditions,
one may identify the profiles. Since the correlation functions
in~(13) and~(14) are the unconnected ones,
the expansion in~(10) now contains the term $n=0$. In the
limit $u_1-u_2\gg1$ the leading contribution is the product of the
diagonal profiles $\phi(v_1)\phi(v_2)$ and the next term contains the
product of the off-diagonal ones, as before. In this way, one obtains
$$\eqalign{
\sigma_{\rm od}^{+-}(v)&\sim\left({L\over\pi}\right)^{-1/8}
\left[\sin\left({\pi v\over L}\right)\right]^{15/8}\cr
\sigma_{\rm od}^{+f}(v)&\sim\left({L\over\pi}\right)^{-1/8}
\left[\sin\left({\pi v\over L}\right)\right]^{7/8}\left[\cos\left({\pi
v\over2L}\right)\right]^{1/2} \tan\left({\pi v\over2L}\right)\cr
\varepsilon_{\rm od}^{+-}(v)&\sim\left({L\over\pi}\right)^{-1}
\sin\left({\pi v\over L}\right)\cos\left({\pi v\over L}\right)\cr
\varepsilon_{\rm od}^{+f}(v)&\sim\left({L\over\pi}\right)^{-1}
\sin\left({\pi v\over L}\right)\; .\cr}  
\eqno(15)
$$
for the off-diagonal order parameter and energy density profiles. The
results of a direct calculation of the off-diagonal 
energy density profiles, in
equations~(A6) and~(A9) of the appendix, are in agreement with the two
last equations of~(15) in the continuum limit.

Diagonal profiles on strips with asymmetric 
boundary conditions for the Ising, 
$Q$-state Potts and $O(N)$ models can be found in
references~[8,10--12] where
they were deduced from the appropriate one-point functions in the
half-plane. 

With mixed boundary conditions the correction term in $\rho_2/\rho_1$
gives a first gap vanishing as $\pi/L$ on the strip: the connection with
surface exponents we had before is lost. Contrary to the
case of symmetric boundary conditions in equations~(4) and~(12),
there is no simple functional form adapted to the different boundary
conditions. 
The diagonal profiles generally contain a regular
leading contribution in their surface finite-size scaling behaviour
whereas the off-diagonal ones behave as $L^{-(x_\phi^{\rm s})_{a,b}}$ 
with $(x_{\sigma}^{\rm s})_{+,-}=(x_{\varepsilon}^{\rm s})_{+,-}=
(x_{\varepsilon}^{\rm s})_f=2$ and $(x_{\sigma}^{\rm s})_f=1/2$. 

One may notice that diagonal and off-diagonal profiles obtained
in~[14] for the random Ising quantum chain are in quite good
agreement with the conformal results given above although the system, which
is strongly anisotropic, is {\it not} conformally invariant. It would be
interesting to check wether this is peculiar to the Ising model or if it
holds true for other $2d$ anisotropic systems as well. 

\Appendix{}
In this appendix, we calculate the off-diagonal energy density profiles on a
strip for the Ising model with different boundary conditions. We work in the
extreme anisotropic limit~[25] where the critical Hamiltonian
reads
$$
{\cal H}=-{\textstyle
1\over2}\left[\sum_{j=1}^{L-1}\sigma_x(j)
\sigma_x(j+1)+\sum_{j=2}^{L-1}\sigma_z(j)
+h_1\sigma_z(1)+h_L\sigma_z(L)\right]\; , 
\eqno(A1)
$$  
where $\sigma_x(j)$ and $\sigma_z(j)$ are Pauli matrices at site $j$. 
The diagonization proceeds in two steps~[26, 27]: first the
Hamiltonian is rewritten as a quadratic form in fermion operators,
$c^\dagger(j)$ and $c(j)$, using the Jordan-Wigner
transformation~[28] and then it is diagonalized
through a canonical transformation to new fermion operators, $\eta_k^\dagger$
and $\eta_k$, such that: 
$$\eqalign{
c(j)&={\textstyle 1\over2}\left\{\sum_{k}[\varphi_k(j)+\psi_k(j)]\eta_k
+[\varphi_k(j)-\psi_k(j)]\eta^\dagger_k\right\}\cr
c^\dagger(j)&={\textstyle
1\over2}\left\{\sum_{k}[\varphi_k(j)+\psi_k(j)]\eta^\dagger_k
+[\varphi_k(j)-\psi_k(j)]\eta_k\right\}\; .\cr}
\eqno(A2)
$$
$\varphi_k$ and $\psi_k$ are normalized eigenvectors, which are conveniently
calculated by a
diagonalization process, as described in~[29].  

With free boundary conditions, $h_1=h_L=1$, the eigenvectors are given by
$$\eqalign{
\varphi_k(j)&=(-1)^j{2\over\sqrt{2L+1}}\cos\left[ {2k-1 \over 2L+1}
\left(j-
1\over2\right)\pi\right]\cr
\psi_k(j)&=(-1)^{j+1}{2\over\sqrt{2L+1}}\sin\left( {2k-1 \over 2L+1}
j\pi\right)\; ,\cr}
\eqno(A3)
$$
with $k=1,2,\dots,L$.
The off-diagonal energy density profile is expressed as
$$
\varepsilon_{\rm od}^{ff}(j)=\langle0\vert\sigma_z(j)\vert\varepsilon\rangle
=2\langle0\vert c^\dagger(j)c(j)\vert\varepsilon\rangle\; ,
\eqno(A4)
$$
where the ground state $\vert0\rangle$ is the fermion vacuum and
$\vert\varepsilon\rangle$ is the lowest eigenstate with two fermionic
excitations so that one obtains: 
$$\fl\eqalign{
\varepsilon_{\rm od}^{ff}(j)&=
\varphi_{1}(j)\psi_{2}(j)-
\varphi_{2}(j)\psi_{1}(j)\cr
&=\!{4\over2L\!+\!1}\!\left\{\sin\!\left({\pi\over2L\!+\!1}\right)
\!\cos\!\left[{4(j\!+\!1/2)\pi\over2L\!+\!1}\right]\!-\!
\cos\!\left({\pi\over2L\!+\!1}\right)
\!\sin\!\left[{(2j\!-\!1/2)\pi\over2L\!+\!1}\right]\right\}.\cr}
\eqno(A5)
$$
When $L\gg1$, the first term may be neglected and the
second one gives the off-diagonal profile in equation~(12) since,
for the $2d$ Ising model, $x_\varepsilon=1$ 
and $x_\varepsilon^{\rm s}=2$ at the
ordinary surface transition. 

When the two boundary spins are fixed one has to put $h_1=h_L=0$ and
then the Hamiltonian in equation~(A1) describes both the $++$ and the $+-$
boundary conditions. Instead of solving this problem explicitly one may
use the duality properties of the quantum Ising model~[25]. Through a
duality transformation, the critical system 
with $++$ ($+-$) boundary condition
is related to the energy (magnetization) sector of the critical free
chain\footnote{\P}{The energy (magnetization) sector corresponds to an
even (odd) number of fermion excitations in the system.}, 
with the correspondence
$L\to L-1$ and $j\to j-1$. Then the  $\varepsilon_{\rm od}^{++}(j)$ energy
density profile is directly given by~(A5) with the above substitutions.

To calculate the $+-$ profile one considers the magnetization
sector of the free chain, where the 
two lowest states are $\eta_1^\dagger \vert 0 \rangle$
and $\eta_2^\dagger \vert 0 \rangle$. Then the $\varepsilon_{\rm od}^{+-}(j)$
profile is given by:
$$\fl\eqalign{
\varepsilon_{\rm od}^{+-}(j)&=
\varphi_{1}(j)\psi_{2}(j)+
\varphi_{2}(j)\psi_{1}(j)\cr
&=\!{16\over2L\!-\!1}\cos\!\left[{(2j\!-\!1)\pi\over4L\!-\!2}\right]
\!\sin\!\left[{(j\!-\!1)\pi\over2L\!-\!1}\right]\left\{
\cos^2\!\left[{(2j\!-\!1)\pi\over4L\!-\!2}\right]-
\!\sin^2\!\left[{(j\!-\!1)\pi\over2L\!-\!1}\right]\right\}.\cr}
\eqno(A6)
$$
which asymptotically behaves as $\varepsilon_{\rm od}^{+-}(j) \approx 2 
\sin(2j\pi/L)/L$ in agreement with the conformal result in~(15).

The $+f$ boundary condition is realized with a vanishing transverse field
$h_1=0$ on the first spin. Then $\sigma_x(1)$ commutes with ${\cal H}$ and
remains fixed in one of its eigenstates corresponding to $\sigma_x(1)=\pm1$.
This introduces a vanishing excitation in the system which doubles the
spectrum. The behaviour of the zero-mode eigenvectors is anomalous:
$$
\varphi_0(j)=0\; ,\qquad\psi_0(j)=(-1)^{j+1}\sqrt{1\over L}\; .
\eqno(A7)
$$
The next fermion state
is characterized by the eigenvectors:
$$\eqalign{
\varphi_1(j)&=(-1)^{j}\sqrt{2\over L}\sin\left[{(j-1)\pi\over L}\right]\cr
\psi_1(j)&=(-1)^{j+1}\sqrt{2\over
L}\cos\left[{(j-1/2)\pi\over L}\right]\; .\cr}  
\eqno(A8)
$$

In terms of eigenvectors the profile is similar to~(A5) and, due to the
vanishing of $\varphi_0(j)$, it takes the form 
$$
\varepsilon_{\rm od}^{+f}(j)=-\varphi_1(j)\psi_0(j)={\sqrt{2}\over
L}\sin\left[{(j-1)\pi\over L}\right]\; , 
\eqno(A9)
$$
in agreement with~(15) in the continuum limit.

The results in equations~(A5), (A6) and~(A9) have been checked
numerically on small-size quantum Ising chains.

\ack We thank H Rieger, M Henkel and D Karevski for helpful
discussions and E~Eisenriegler for a critical reading of the
manuscript. FI is indebted to the Hungarian National Research
Fund for financial support under grants no OTKA TO12830, 17485 and 23642.

\references
\numrefjl{[1]}{Fisher M E and de Gennes P G 1978}{C. R. Acad. Sci. 
Paris {\rm B}}{287}{207}
\numrefbk{[2]}{Binder K 1983}{Phase Transitions and Critical
Phenomena}{vol~8 ed C~Domb and J~L~Lebowitz 
(London: Academic Press) p~1}
\numrefbk{[3]} {Diehl H W 1986}{Phase Transitions and Critical
Phenomena}{vol~10 ed C~Domb and J~L~Lebowitz 
(London: Academic Press) p~75}
\numrefjl{[4]}{Au-Yang H and Fisher M E 1980}{\PR\ {\rm B}}{21}{3956}
\numrefjl{[5]}{Bariev R Z 1988}{Theor. Math. Phys.}{77}{1090}
\numrefjl{[6]}{Burkhardt T W and Eisenriegler E 
1985}{\JPA}{18}{L83}
\numrefjl{[7]}{Burkhardt T W and Cardy J L 1987}{\JPA}{20}{L233}
\numrefjl{[8]}{Burkhardt T W and Guim I 1987}{\PR\ {\rm B}}{36}{2080}  
\numrefjl{[9]}{Cardy J L 1990}{\PRL}{65}{1443}
\numrefjl{[10]}{Burkhardt T W and Xue T 1991}{\PRL}{66}{895} 
\numrefjl{[11]}{Burkhardt T W and Xue T 1991}{Nucl. Phys. {\rm B}}{354}{653}
\numrefjl{[12]}{Burkhardt T W and Eisenriegler E 1994}{Nucl. Phys. {\rm
B}}{424}{487}  
\numrefjl{[13]}{Karevski D and Henkel M 1997}{\PR\ {\rm
B}}{55}{6429}   
\numrefjl{[14]}{Igl\'oi F anf Rieger H 1997}{\PRL}{78}{2473}
\numrefjl{[15]}{Mikheev L V and Fisher M E 1994}{\PR\  {\rm B}}{49}{378} 
\numrefjl{[16]}{Czerner P and Ritschel 
U 1996}{\PRL}{77}{3645} \ \ \ \ \ \ \ \ 
and {\it Essen preprints} (cond-mat/9609120, cond-mat/9609140)  
\numrefjl{[17]}{Cardy J L 1983}{\JPA}{17}{L385}  
\numrefbk{[18]}{Cardy J L 1987}{Phase Transitions
and  Critical Phenomena}{vol~11 ed C~Domb and J~L~Lebowitz 
(London: Academic Press) p~55}
\numrefjl{[19]}{Bray A J and Moore M A 1977}{\JPA}{10}{1927}
\numrefjl{[20]}{Diehl H W and Smock M 1993}{\PR\ {\rm B}}{47}{5841} 
\numrefjl{[21]}{Burkhardt T W and Diehl H W 1994}{\PR\ {\rm B}}{50}{3894} 
\numrefjl{[22]}{Guim I and Burkhardt TW 1989}{\JPA}{22}{1131}
\numrefjl{[23]}{Eisenriegler E, Krech M and Dietrich S 1996}{\PR\
{\rm B}}{53}{14377}
\numrefjl{[24]}{Cardy J L 1984}{Nucl. Phys. {\rm B}}{240}{514} 
\numrefjl{[25]}{Kogut J B 1979}{\RMP}{51}{659}
\numrefjl{[26]}{Lieb E H, Schultz T D and Mattis D C 1961}{\APNY}{16}{406}
\numrefjl{[27]}{Pfeuty P 1979}{Ann. Phys.,Paris}{\bf 57}{79}
\numrefjl{[28]}{Jordan P and Wigner E 1928}{\ZP}{47}{631}
\numrefjl{[29]}{Igl\'oi F and Turban L 1996}{\PRL}{\bf 77}{1206}
\vfill\eject\bye